\documentclass{llncs}
\usepackage{amsmath}
\usepackage{amssymb}
\usepackage{algpseudocode}
\usepackage{graphicx}

\newif\iflong
\longtrue

\newcommand{\lvec}[1]{\overset{{}_{\leftarrow}}{#1}}
\newcommand{\lrange}[1]{\overleftarrow{#1}}
\newcommand{\BWT}{\mathit{BWT}}
\newcommand{\SA}{\mathit{SA}}
\newcommand{\lab}{\mathit{lab}}
\newcommand{\lcp}{\mathit{lcp}}
\newcommand{\rlcp}{\mathit{rlcp}}
\newcommand{\mycap}[1]{\textbf{\textit{#1}}}

\algtext*{EndFunction}
\algtext*{EndProcedure}
\algtext*{EndFor}
\algtext*{EndWhile}
\algtext*{EndIf}
\algnewcommand{\IIf}[1]{\State\algorithmicif\ #1\ \algorithmicthen\ }
\algnewcommand{\EndIIf}{\unskip}
\algnewcommand{\WWhile}[1]{\State\algorithmicwhile\ #1\ \algorithmicdo\ }
\algnewcommand{\EndWWhile}{\unskip}
\algloopdefx{NIf}[1]{\textbf{if} #1}
\algloopdefx{NElse}{\textbf{else}}
\algloopdefx{NElseIf}{\textbf{else if}}
\algloopdefx{NForAll}[1]{\textbf{for each} #1 \textbf{do}}
\algloopdefx{NWhile}[1]{\textbf{while} #1 \textbf{do}}
\algloopdefx{NFor}[1]{\textbf{for} #1}

\title {Faster Lightweight Lempel-Ziv Parsing}

\author{Dmitry Kosolobov}
\institute{Ural Federal University, Ekaterinburg, Russia\\ \email{dkosolobov@mail.ru}}

\begin{document}

\maketitle

\begin{abstract}
We present an algorithm that computes the Lempel-Ziv decomposition in $O(n(\log\sigma + \log\log n))$ time and $n\log\sigma + \epsilon n$ bits of space, where $\epsilon$ is a constant rational parameter, $n$ is the length of the input string, and $\sigma$ is the alphabet size. The $n\log\sigma$ bits in the space bound are for the input string itself which is treated as read-only.
\end{abstract}

\section{Introduction}

The Lempel-Ziv decomposition~\cite{LempelZiv} is a basic technique for data compression and plays an important role in string processing. It has several modifications used in various compression schemes. The decomposition considered in this paper is used in LZ77-based compression methods and in several compressed text indexes designed to efficiently store and search massive highly-repetitive data sets.

The standard algorithms computing the Lempel-Ziv decomposition work in $O(n\log\sigma)$\footnote{Throughout the paper, $\log$ denotes the logarithm with the base~$2$.} time and $O(n\log n)$ bits of space, where $n$ is the length of the input string and $\sigma$ is the alphabet size. It is known that this is the best possible time for the general alphabets~\cite{Kosolobov}. However, for the most important case of integer alphabet, there exist algorithms working in $O(n)$ time and $O(n\log n)$ bits (see \cite{FischerIKoppl} for references). When $\sigma$ is small, this number of bits is too big compared to the $n\log\sigma$ bits of the input string and can be prohibitive. To address this issue, several algorithms using $O(n\log\sigma)$ bits were designed.

The main contribution of this paper is a new algorithm computing the Lempel-Ziv decomposition in $O(n(\log\sigma + \log\log n))$ time and $n\log\sigma + \epsilon n$ bits of space, where $\epsilon$ is a constant rational parameter. The $n\log\sigma$ bits in the space bound are for the input string itself which is treated as read-only. The following table lists the time and space required by existing approaches to the Lempel-Ziv parsing in $O(n\log\sigma)$ bits of space.

\begin{tabular}{|c|c|c|l|}
\hline
Time                       & Bits of space       & Note & Author(s) \\
\hline
$O(n\log\sigma)$          & $O(n\log\sigma)$     &  &Ohlebusch and Gog \cite{OhlebuschGog} \\
$O(n\log^3 n)$            & $n\log\sigma + O(n)$ & online & Okanohara and Sadakane \cite{OkanoharaSadakane} \\
$O(n\log^2 n)$            & $O(n\log\sigma)$     & online & Starikovskaya \cite{Starikovskaya} \\
$O(n\log n)$               & $O(n\log\sigma)$    & online & Yamamoto et al. \cite{YamamotoIBannaiInenagaTakeda} \\
$O(n\log n\log\log\sigma)$ & $n\log\sigma+\epsilon n$ & & K\"arkk\"ainen et al. \cite{KarkkainenKempaPuglisi}\\
$O(n(\log\sigma + \log\log n))$ & $n\log\sigma + \epsilon n$ & & this paper\\
\hline
\end{tabular}

By a more careful analysis, one can show that when $\epsilon$ is not a constant, the running time of our algorithm is $O(\frac{n}{\epsilon}(\log\sigma + \log\frac{\log n}{\epsilon}))$; we omit the details here.

\subsubsection{Preliminaries.}Let $w$ be a string of length $n$. Denote $|w| = n$. We write $w[0], w[1], \ldots, w[n{-}1]$ for the letters of $w$ and $w[i..j]$ for $w[i]w[i{+}1]\cdots w[j]$. A string can be \emph{reversed} to get $\lvec{w} = w[n{-}1]\cdots w[1]w[0]$ called the \emph{reversed $w$}. A string $u$ is a \emph{substring} (or \emph{factor}) of $w$ if $u=w[i..j]$ for some $i$ and $j$. The pair $(i,j)$ is not necessarily unique; we say that $i$ specifies an \emph{occurrence} of $u$ in $w$. A string can have many occurrences in another string. For $i,j \in \mathbb{Z}$, the set $\{k\in \mathbb{Z} \colon i \le k \le j\}$ is denoted by $[i..j]$; $[i..j)$ denotes $[i..j{-}1]$.

Throughout the paper, $s$ denotes the input string of length $n$ over the integer alphabet $[0..\sigma)$. Without loss of generality, we assume that $\sigma \le n$ and $\sigma$ is a power of two. Thus, $s$ occupies $n\log\sigma$ bits. Simplifying the presentation, we suppose that $s[0]$ is a special letter that is smaller than any letter in $s[1..n{-}1]$.

Our model of computation is the unit cost word RAM with the machine word size at least $\log n$ bits. Denote $r = \log_\sigma n = \frac{\log n}{\log\sigma}$. For simplicity, we assume that $\log n$ is divisible by $\log\sigma$. Thus, one machine word can contain a string of length $\le r$; we say that it is a \emph{packed string}. Any substring of $s$ of length $r$ can be packed in a machine word in constant time by standard bitwise operations. Therefore, one can compare any two substrings of $s$ of length $k$ in $O(k/r + 1)$ time.

The \emph{Lempel-Ziv decomposition of $s$} is the decomposition $s = z_1z_2\cdots z_l$ such that each $z_i$ is either a letter that does not occur in $z_1z_2\cdots z_{i-1}$ or the longest substring that occurs at least twice in $z_1z_2\cdots z_i$ (e.g., $s = a\cdot b\cdot b\cdot abbabb\cdot c\cdot ab\cdot ab$). The substrings $z_1, z_2, \ldots, z_l$ are called the \emph{Lempel-Ziv factors}. Our algorithm consecutively reports the factors in the form of pairs $(|z_i|, p_i)$, where $p_i$ is either the position of a nontrivial occurrence of $z_i$ in $z_1z_2\cdots z_i$ (it is called an \emph{earlier occurrence of $z_i$}) or $z_i$ itself if $z_i$ is a letter that does not occur in $z_1z_2\cdots z_{i-1}$. The reported pairs are not stored in main memory.

Fix a rational constant $\epsilon > 0$. It suffices to prove that our algorithm works in $O(n(\log\sigma + \log\log n))$ time and $n\log\sigma + O(\epsilon n)$ bits: the substitution $\epsilon' = c\epsilon$, where $c$ is the constant under the bit-$O$, gives the required $n\log\sigma + \epsilon'n$ bits with the same working time. We use different approaches to process the Lempel-Ziv factors of different lengths. In Section~\ref{SectShortFactors} we show how to process ``short'' factors of length ${<}r/2$. In Section~\ref{SectMidFactors} we describe new compact data structures that allow us to find all ``medium'' factors of length ${<}(\log n/\epsilon)^2$. In Section~\ref{SectLongFactors} we apply the clever technique of~\cite{BurkhardtKarkkainen} for the analysis of all other ``long'' factors.

\section{Short Factors}\label{SectShortFactors}

In this section we consider the Lempel-Ziv factors of length $<r/2$, so we assume $r \ge 2$. Suppose the algorithm has reported the factors $z_1, z_2, \ldots, z_{k-1}$ and now we process $z_k$. Denote $p = |z_1z_2\cdots z_{k-1}|$. We maintain arrays $H_1, H_2, \ldots, H_{\lceil r/2\rceil}$ defined as follows: for $i \in [1..\lceil\frac{r}{2}\rceil]$, the array $H_i$ contains $\sigma^i$ integers such that for any $x \in [0..\sigma^i)$, either $H_i[x]$ equals the position from $[0..p)$ of an occurrence in $s$ of the packed string $x$ of length $i$ or $H_i[x] = -1$ if there are no such positions.

For each $i\in [1..r]$ and $j\in [0..n]$, denote by $x^j_i$ the packed string $s[j..j{+}i{-}1]$. We have $H_1[x^p_1] = -1$ iff $z_k$ is a letter that does not appear in $s[0..p{-}1]$; in this case the algorithm reports $z_k$ immediately. Further, we have $H_{\lceil r/2\rceil}[x^p_{\lceil r/2\rceil}] \ne -1$ iff $|z_k| \ge \frac{r}2$; this case is considered in Sections~\ref{SectMidFactors},~\ref{SectLongFactors}. Suppose $H_1[x^p_1] \ne -1$ and $H_{\lceil r/2\rceil}[x^p_{\lceil r/2\rceil}] = -1$. Our algorithm finds the minimal $q \in [2..\lceil \frac{r}2\rceil]$ such that $H_q[x^p_q]=-1$. Then we obviously have $|z_k| = q{-}1$ and $H_{|z_k|}[x^p_{|z_k|}]$ is the position of an earlier occurrence of $z_k$. Clearly, the algorithm works in $O(|z_k|)$ time.

The inequality $r = {\log n}/{\log\sigma} \ge 2$ implies $\sigma \le \sqrt{n}$. Thus, $H_1, H_2, \ldots, H_{\lceil r/2\rceil}$ altogether occupy at most $\sigma^{\lceil r/2\rceil}r\log n \le \sigma^{\frac{r}2}\sigma^{\frac{1}2} r\log n \le n^{\frac{3}{4}}r\log n = o(n)$ bits.

To maintain $H_1, \ldots, H_{\lceil r/2\rceil}$, we consecutively examine the positions $j = 0, 1, \ldots, p{-}1$ and for those positions, for which $H_{\lceil r/2\rceil}[x^j_{\lceil r/2\rceil}] = -1$, we perform the assignments $H_1[x^j_1] \gets j, H_2[x^j_2] \gets j, \ldots, H_{\lceil r/2\rceil}[x^j_{\lceil r/2\rceil}] \gets j$. Hence, we execute these assignments for at most $\sigma^{\lceil r/2\rceil}$ positions and the overall time required for the maintenance of $H_1, \ldots, H_{\lceil r/2\rceil}$ is $O(n + r\sigma^{\lceil r/2\rceil}) = O(n)$.

\section{Medium Factors}\label{SectMidFactors}

Suppose the algorithm has reported the Lempel-Ziv factors $z_1, z_2, \ldots, z_{k-1}$ and already decided that $|z_k| \ge \frac{r}2$ applying the procedure of Section~\ref{SectShortFactors}. Denote $p = |z_1z_2\cdots z_{k-1}|$, $\tau = \lceil\frac{\log n}{\epsilon}\rceil$, and $b = \lceil\epsilon n / (\log\sigma + \log\log n)\rceil$. We assume $p{+}b{+}\tau^2 < n$; the case $p{+}b{+}\tau^2 \ge n$ is analogous. Our algorithm processes $s[0..p{+}b]$ and reports not only $z_k$ but also all Lempel-Ziv factors starting in positions $[p..p{+}b]$.

The algorithm consists of three phases: the first one builds for other phases an indexing data structure on the string $s[p..p{+}b]$ in $O(b\log\sigma)$ time and $O(b(\log\sigma + \log\log n)) = O(\epsilon n)$ bits; the second phase scans $s[0..p{+}b]$ in $O(n)$ time and fills a bit array $\mathit{lz}[0..b]$ so that for any $i\in [0..b]$, $\mathit{lz}[i] = 1$ iff there is a Lempel-Ziv factor starting in the position $p{+}i$; finally, the last phase scans $s[0..p{+}b]$ in $O(n)$ time and reports earlier occurrences of the found Lempel-Ziv factors. Thus, the overall time required by this algorithm is $O((n + b\log\sigma)\frac{n}b) = O(n(\log\sigma + \log\log n))$.

The data structures we use can search only the Lempel-Ziv factors of length $<\tau^2$; we delegate the longer factors to the procedure of Section~\ref{SectLongFactors}. This restriction allows us to make our structures fast and compact. More precisely, our algorithm consecutively computes the lengths of the Lempel-Ziv factors starting in $[p..p{+}b]$ and once we have found a factor of length $\ge \tau^2$, we invoke the procedure of Section~\ref{SectLongFactors} to compute the length and an earlier occurrence of this factor.

\subsection{Main Tools}

Let $x$ be a string of length $d{+}1$. Denote $\lvec{x}_i = \lrange{x[0..i]}$. The \emph{suffix array of $\lvec{x}$} is the permutation $\SA[0..d]$ of the integers $[0..d]$ such that $\lvec{x}_{\SA[0]} < \lvec{x}_{\SA[1]} < \ldots < \lvec{x}_{\SA[d]}$ in the lexicographical order. The \emph{Burrows-Wheeler transform}~\cite{BurrowsWheeler} of $\lvec{x}$ is the string $\BWT[0..d]$ such that $\BWT[i] = x[\SA[i]{+}1]$ if $\SA[i] < d$ and $\BWT[i] = x[0]$ otherwise. We equip $\BWT$ with the function $\Psi$ defined as follows: $\Psi(i) = \SA^{-1}[\SA[i] + 1]$ if $\SA[i] < d$ and $\Psi(i) = 0$ otherwise.
\begin{lemma}[see~\cite{HonSadakaneSung}]
The string $\BWT$ and the function $\Psi$ for a string $\lvec{x}$ of length $d{+}1$ over the alphabet $[0..\sigma)$ can be constructed in $O(d\log\log\sigma)$ time and $O(d\log\sigma)$ bits of space; $\Psi$ is encoded in $O(d\log\sigma)$ bits with $O(1)$ access time.\label{BWT}
\end{lemma}
\iflong
\begin{example}
Consider the string $x = \$aabadcaababadcaaba$.
$$\scriptsize{
\begin{array}{r|c|c|c|c}
x[0..\SA[i]] & \BWT[i] & \SA[i] & \Psi(i) & i \\
\hline
\$ & a & 0 & 1 & 0 \\
\$a & a & 1 & 2 & 1 \\
\$aa & b & 2 & 11 & 2 \\
\$aabadcaa & b & 8 & 12 & 3 \\
\$aabadcaababadcaa & b & 16 & 13 & 4 \\
\$aaba & d & 4 & 17 & 5 \\
\$aabadcaaba & b & 10 & 14 & 6 \\
\$aabadcaababadcaaba & \$ & 18 & 0 & 7 \\
\$aabadcaababa & d & 12 & 18 & 8 \\
\$aabadca & a & 7 & 3 & 9 \\
\$aabadcaababadca & a & 15 & 4 & 10 \\
\$aab & a & 3 & 5 & 11 \\
\$aabadcaab & a & 9 & 6 & 12 \\
\$aabadcaababadcaab & a & 17 & 7 & 13 \\
\$aabadcaabab & a & 11 & 8 & 14 \\
\$aabadc & a & 6 & 9 & 15 \\
\$aabadcaababadc & a & 14 & 10 & 16 \\
\$aabad & c & 5 & 15 & 17 \\
\$aabadcaababad & c & 13 & 16 & 18
\end{array}
}$$
\end{example}
\fi

In the \emph{dynamic weighted ancestor (WA for short) problem} one has 1) a weighted tree, where the weight of each vertex is greater than the weight of parent, 2) the queries finding for a vertex $v$ and number $i$ the ancestor of $v$ with the minimal weight $\ge i$, 3) the updates inserting new vertices. Let $v$ be a vertex of a trie $T$ ($v \in T$ for short). Denote by $\lab(v)$ the string written on the path from the root to $v$. We treat tries as weighted trees: $|\lab(v)|$ is the weight of $v$.
\begin{lemma}[see~\cite{KopelowitzLewenstein}]
For a weighted tree with at most $k$ vertices, the dynamic WA problem can be solved in $O(k\log k)$ bits of space with queries and updates working in $O(\log k)$ amortized time.\label{WeightedAncestor}
\end{lemma}
One can easily modify the proof of~\cite{KopelowitzLewenstein} for a special case of this problem when the weights are integers $[0..\tau^2]$ and the height of the tree is bounded by $\tau^2$.
\iflong
\begin{lemma}
\else
\begin{lemma}[see~\cite{KopelowitzLewenstein}]
\fi
Let $T$ be a weighted tree with at most $m \le n$ vertices, the weights $[0..\tau^2]$, and the height ${\le}\tau^2$. The dynamic WA problem for $T$ can be solved in $O(m(\log m + \log\log n))$ bits of space with queries and updates working in $O(1)$ amortized time using a shared table of size $o(n)$ bits.\label{WeightedAncestorSpec}
\end{lemma}
\newcommand{\WASpecProof}{
\begin{proof}
In~\cite{KopelowitzLewenstein}, using $O(m\log m)$ additional bits of space, the general problem for a tree with $m$ vertices, the weights $[0..\tau^2]$, and the height $\le\tau^2$ is reduced to the same problem for subtrees with at most $\log\log m$ vertices and the problem of the maintenance of a set of dynamic predecessor data structures on the weights $[0..\tau^2]$ so that each of these predecessor structures contains at most $\tau^2$ weights and all they contain $O(m)$ weights in total. Each query or update on the tree requires a constant number of queries/updates on the subtrees of size $\le \log\log m$ and on the predecessor structures.

Since the weights are bounded by $\tau^2$, a subtree with at most $\log\log m$ vertices fits in $O(\log\log m\log\tau) = O((\log\log n)^2)$ bits. So, we can perform queries and updates on these trees in $O(1)$ time using a shared table of size $O(2^{O((\log\log n)^2)}\log^{O(1)} n) = o(n)$ bits. Further, one can organize a dynamic predecessor data structure with at most $\tau^2$ elements as a $B$-tree of a constant depth with $O(\sqrt{\tau})$-element predecessor structures on each level. Any predecessor structure with $O(\sqrt{\tau})$ weights fits in $O(\sqrt{\tau}\log\log n)$ bits and therefore, one can perform all operations on these small structures with the aid of a shared table of size $O(2^{\sqrt{\tau}\log\log n}\log^{O(1)} n) = o(n)$ bits. Thus we can perform all operations on the source predecessor structure in $O(1)$ time.
\qed
\end{proof}
}
\iflong
\WASpecProof
\fi

Denote by $\lcp(t_1, t_2)$ the length of the longest common prefix of the strings $t_1$ and $t_2$. Denote $\rlcp(i, j) = \min\{\tau^2, \lcp(x'_{\SA[i]}, x'_{\SA[j]})\}$.
\begin{lemma}[see~\cite{BellerGogOhlebuschSchnattinger}]
For a string $x$ of length $d{+}1$, using $\BWT$ of $\lvec{x}$, one can compute an array $\rlcp[0..d{-}1]$ such that $\rlcp[i] = \rlcp(i,i{+}1)$, for $i\in[0..d)$, in $O(d\log\sigma)$ time and $O(d\log\sigma)$ bits; the array occupies $O(d\log\log n)$ bits.\label{LCP}%
\end{lemma}

\subsection{Indexing Data Structure}\label{SubsectConstruct}

\noindent\mycap{Trie.} Denote $d = 1{+}b{+}\tau^2$. The algorithm creates a string $x$ of length $d{+}1$ and copies the string $s[p..p{+}b{+}\tau^2]$ in $x[1..d]$; $x[0]$ is set to a special letter less than any letter in $x[1..d]$. Let $\SA$ be the suffix array of $\lvec{x}$ (we use it only conceptually). Denote $x'_i = \lrange{x[i{-}\tau^2{+}1..i]}$ (we assume that $x[-1], x[-2], \ldots$ are equal to $x[0]$). Here we discuss the design of our indexing data structure, a carefully packed in $O(d(\log\sigma + \log\log n))$ bits augmented compact trie of the strings $x'_0, x'_1, \ldots, x'_d$.

For simplicity, suppose $d$ is a multiple of $r$.  The skeleton of our structure is a compact trie $Q_0$ of the strings $\{x'_{\SA[jr]} \colon j\in [0..d/r]\}$. We augment $Q_0$ with the WA structure of Lemma~\ref{WeightedAncestorSpec}. Each vertex $v \in Q_0$ contains the following fields: 1) the pointer to the parent of $v$ (if any); 2) the pointers to the children of $v$ in the lexicographical order; 3) the length of $\lab(v)$; 4) the length of the string written on the edge connecting $v$ to its parent (if any).

Notice that the fields 3)--4) fit in $O(\log\log n)$ bits. Clearly, $Q_0$ occupies $O((d/r)\log n) = O(d\log\sigma)$ bits of space. The pointers to the substrings of $x$ written on the edges of $Q_0$ are not stored, so, one cannot use $Q_0$ for searching.%
\begin{figure}
\center
\includegraphics[scale=0.35]{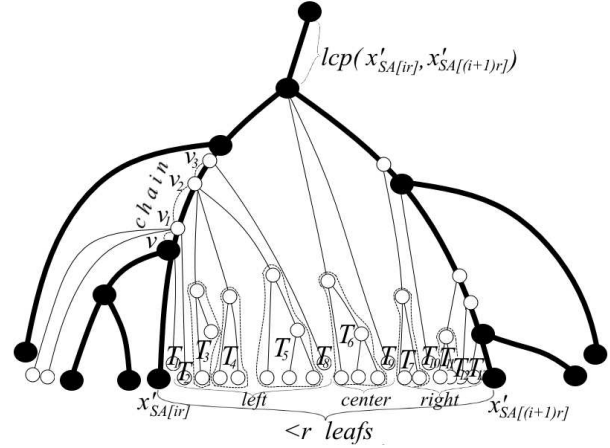}
\caption{\small Solid vertices and edges are from $Q_0$.}
\label{fig:trieconstruct}
\end{figure}

We create an array $L[0..d/r]$ such that for $i\in[0..d/r]$, $L[i]$ is the pointer to the leaf of $Q_0$ corresponding to $x'_{\SA[ir]}$. Now we build a compact trie $Q$ inserting the strings $\{x'_{\SA[ir{+}j]}\}_{j=1}^{r{-}1}$ in $Q_0$ for each $i \in [0..d/r)$ as follows. For a fixed $i$, these strings add to $Q_0$ trees $T_1, \ldots, T_l$ attached to the branches $x'_{\SA[ir]}$ and $x'_{\SA[(i{+}1)r]}$ in $Q_0$ (see Fig.~\ref{fig:trieconstruct}). We store $T_1, \ldots, T_l$ in a contiguous memory block $F_i$. The pointer to $F_i$ is stored in the leaf of $Q_0$ corresponding to $x'_{\SA[ir]}$, so, one can find $F_i$ in $O(1)$ time using $L$. Since $T_1, \ldots, T_l$ have at most $2r$ vertices in total, $O(\log\log n)$ bits per vertex suffice for the fields 1)--4). Now we discuss how $T_1, \ldots, T_l$ are attached to $Q_0$. Consider $v \in Q_0$ and the vertices $v_1,\ldots,v_h$ splitting the edge connecting $v$ to its parent in $Q_0$. Let $T_{i_1}, \ldots, T_{i_g}$ be the trees that must be attached to $v, v_1,\ldots, v_h$ (see Fig.~\ref{fig:trieconstruct}). We add to $v$ a memory block $N_v$ containing the WA structure of Lemma~\ref{WeightedAncestorSpec} for the chain $v, v_1, \ldots, v_h$ with the weights $|\lab(v)|, |\lab(v_1)|, \ldots, |\lab(v_h)|$. Each of the vertices $v,v_1,\ldots,v_h$ in this chain contains the $O(\log\log n)$-bit pointers (inside $F_i$) to the roots of $T_{i_1}, \ldots, T_{i_g}$ attached to this vertex. Hence, $N_v$ occupies $O((h + g)\log\log n)$ bits. One can find the children for each of the vertices $v, v_1,\ldots, v_h$ in $O(1)$ time using $Q_0$ and the chain in the block $N_v$. Further, one can find, for any $j \in [1..g]$, the parent of the root of $T_{i_j}$ in $O(1)$ time by a WA query on $Q_0$ to find a suitable $v$ and a WA query on the chain in $N_v$. Finally, we augment each $T_i$ with the WA structure of Lemma~\ref{WeightedAncestorSpec}. Thus, by Lemma~\ref{WeightedAncestorSpec}, $T_1, \ldots, T_l$ add at most $O(r\log\log n)$ bits to $Q$.

For each $i \in [0..d/r)$, we augment the leaf referred by $L[i]$ with an array $L_i[0..r{-}2]$ such that for $j \in [0..r{-}2]$, $L_i[j]$ is the $O(\log\log n)$-bit pointer (inside $F_i$) to the leaf of $Q$ corresponding to $x'_{\SA[ir{+}1{+}j]}$. So, for any $j \in [0..d]$, one can easily find the leaf of $Q$ corresponding to $x'_{\SA[j]}$ in $O(1)$ time via $L$ and $L_{\lfloor j/r\rfloor}$. Finally, the whole described structure $Q$ occupies $O(d(\log\sigma + \log\log n))$ bits.

\noindent\mycap{Prefix links.} Consider $v \in Q$. Denote by $[i_v..j_v]$ the longest segment such that for each $i\in [i_v..j_v]$, $x'_{\SA[i]}$ starts with $\lab(v)$ (see Fig.~\ref{fig:bwttrie}). Let $\BWT$ be the Burrows-Wheeler transform of $\lvec{x}$. Denote the set of the letters of $\BWT[i_v..j_v]$ by $P_v$. We associate with $v$ the \emph{prefix links} mapping each $c \in P_v$ to an integer $p_v(c) \in [i_v..j_v]$ such that $x[\SA[p_v(c)]{+}1] = c$ (there might be many such $p_v(c)$; we choose any). The prefix links correspond to the well-known \emph{Weiner-links}. Hence, $Q$ has at most $O(d)$ prefix links. Observe that $P_u \supset P_v$ for any ancestor $u$ of $v$. The problem is to store the prefix links in $O(d(\log\sigma + \log\log n))$ bits.
\begin{figure}
\center
\includegraphics[scale=0.40]{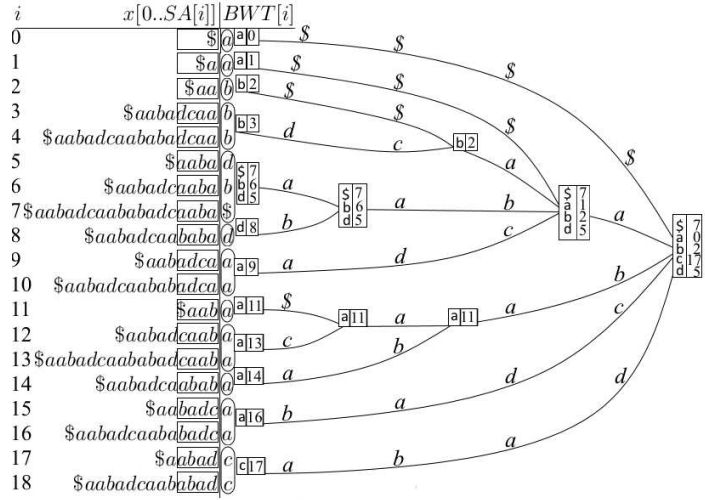}
\caption{\small $\tau^2 = 4$, the prefix links associated with vertices are in squares.}
\label{fig:bwttrie}
\end{figure}

Fix $i \in [0..d)$. Denote by $V_i$ the set of the vertices $v \notin Q_0$ such that $v$ does not have descendants from $Q_0$ and lies between branches $x'_{\SA[ir]}$ and $x'_{\SA[(i{+}1)r]}$. We associate with each $v \in V_i$ a dictionary $D_v$ mapping each $c \in P_v$ to $p_v(c){-}ir$ and store all $D_v$, for $v\in V_i$, in a contiguous memory block $H_i$. Since $|V_i| < r$ and $P_v$ is a subset of $\BWT[ir..(i{+}1)r]$, we have $p_v(c){-}ir \in [1..r)$ and all $D_v$, for $v \in V_i$, occupy overall $O(\sum_{v\in V_i}|P_v|(\log\sigma + \log\log n)) = O(r^2(\log\sigma + \log\log n))$ bits of space. Therefore, we can store in each $v\in V_i$ the $O(\log\log n)$-bit pointer to $D_v$ (inside $H_i$). The pointer to $H_i$ itself is stored in the leaf referred by $L[i]$.%

Consider $v \notin Q_0$ such that $v$ lies on an edge connecting a vertex $w \in Q_0$ to its parent in $Q_0$. Let $x'_{\SA[j_1r]}$ and $x'_{\SA[j_2r]}$ be the strings corresponding to the leftmost and rightmost descendant leaves of $w$ contained in $Q_0$. We split $P_v$ on three subsets: $P_1 = \{c\in P_v \colon p_v(c) < j_1r\}$, $P_2 = \{c\in P_v \colon p_v(c) > j_2r\}$, $P_3 = P_v \setminus (P_1\cup P_2)$. Clearly $P_3 \subset P_w\subset P_v$. Hence, we can use $P_w$ instead of $P_3$ and store only the sets $P_1$ and $P_2$ in a way similar to that discussed above.

Suppose $v\in Q_0$. Let for $c\in P_v$, $j_c \in [i_v..j_v]$ be the position of the first occurrence of $c$ in $\BWT[i_v..j_v]$. Clearly, we can set $p_v(c) = j_c$. We add to $v$ a dictionary mapping each $c\in P_v$ to $h_c = |\{c' \in P_v \colon j_{c'} < j_c\}|$. Denote $q = |P_v|$. Since $q \le \sigma$, the dictionary occupies $O(q\log\sigma)$ bits. Now it suffices to map $h_c$ to $j_c$. Let $j'_0, \ldots, j'_{q-1}$ denote all $j_c$, for $c\in P_v$, in increasing order. Obviously $j'_{h_c} = j_c$. The idea is to sample each $(\tau^2\log n)$th position in $\BWT$. We add to $v$ a bit array $A_v[0..q{-}1]$ indicating the sampled $j'_0,\ldots,j'_{q-1}$: $A_v[0] = 1$ and for $h \in [1..q)$, $A_v[h] = 1$ iff $j'_{h{-}1} < l\tau^2\log n \le j'_h$ for an integer $l$; $A_v$ is equipped with the structure of~\cite{RamanRamanRao} supporting the queries $\mathrm{r}_{A_v}(h) = \sum_{i=0}^h A_v[i]$ in $O(1)$ time and $o(q)$ additional bits. The sampled sequence $\{j'_h \colon A_v[h] = 1\}$ is stored in an array $B_v$. Finally, we add an array $C_v[0..q{-}1]$ such that $C_v[h] = j'_h - B_v[\mathrm{r}_{A_v}(h){-}1]$. Now we map $h$ to $j'_h$ as follows: $j'_h = B_v[\mathrm{r}_{A_v}(h){-}1] + C_v[h]$. Clearly, each value of $C_v$ is in the range  $[0..\tau^2\log n]$ and hence, $C_v$ occupies $O(q\log(\tau^2\log n)) = O(q\log\log n)$ bits. It suffices to estimate the space consumed by $B_v$. Since the number of the vertices in $Q_0$ is $O(d/r)$ and the height of $Q$ is at most $\tau^2$, all $B_v$ arrays occupy at most $O((d/r)\log n + \frac{d}{\tau^2\log n} \tau^2\log n) = O(d\log\sigma)$ bits in total.

\noindent\mycap{Construction of $Q$.} Initially, $Q$ contains one leaf corresponding to $x'_{\SA[0]}$. We consecutively insert $x'_{\SA[1]}, \ldots, x'_{\SA[d]}$ in $Q$ in groups of $r$ elements. During the construction, we maintain on $Q$ a set of the dynamic WA structures of Lemma~\ref{WeightedAncestorSpec} in such a way that one can answer any WA query on $Q$ in $O(1)$ time.

Suppose we have inserted $x'_{\SA[0]}, \ldots, x'_{\SA[ir]}$ in $Q$ and now we are to insert $x'_{\SA[ir{+}1]}, \ldots, x'_{\SA[(i{+}1)r]}$. We first allocate the memory block $F_i$ required for new vertices. Using Lemma~\ref{LCP}, we compute $\rlcp(j{-}1, j)$ for all $j \in [ir{+}1..(i{+}1)r]$. Since $\rlcp(j_1, j_2) = \min\{\rlcp(j_1, j_2{-}1), \rlcp(j_2{-}1, j_2)\}$, the algorithm can compute $\rlcp(ir, ir{+}j)$ for all $j\in[1..r]$ in $O(r)$ time. Using the WA query on the leaf $x'_{\SA[ir]}$ and the value $\rlcp(ir, (i{+}1)r)$, we find the position where we insert a new leaf $x'_{\SA[(i{+}1)r]}$. Similarly, using the WA queries, we consecutively insert $x'_{\SA[ir{+}j]}$ for $j = 1,2,\ldots$ as long as $\rlcp(ir, ir{+}j) > \rlcp(ir, (i{+}1)r)$ and then all other $x'_{\SA[(i{+}1)r{-}j]}$ for $j=1,2,\ldots$ (Fig.~\ref{fig:trieconstruct}). All related WA structures, the arrays $L, L_i$, the pointers, and the fields for the vertices are built in an obvious way.

One can construct the prefix links of a vertex from those of its children in $O(q\log\sigma)$ time, where $q$ is the number of the links in the children. As there are at most $O(d)$ prefix links, one DFS traverse of $Q$ builds them in $O(d\log\sigma)$ time.

Finally, using the result of~\cite{HagerupMiltersenPagh}, the algorithm converts in $O(d\log\sigma)$ time all dictionaries in the prefix links of the resulting trie $Q$ in the perfect hashes with $O(1)$ access time. So, one can access any prefix link in $O(1)$ time.

\subsection{Algorithm for Medium Factors}

In the \emph{dynamic marked descendant problem} one has a tree, a set of marked vertices, the queries asking whether there is a marked descendant of a given vertex, and the updates marking a given vertex. We assume that each vertex is a descendant of itself. We solve this problem on $Q$ as follows\iflong.\fi%
\begin{lemma}
In $O(d(\log\sigma + \log\log n))$ bits one can solve the dynamic marked descendant problem on $Q$ so that any $k$ queries and updates take $O(k + d)$ time.\label{MarkedDescendant}
\end{lemma}
\newcommand{\MarkedDescProof}{
\begin{proof}
Let $q$ be the number of the vertices in $Q$. Obviously $q = O(d)$. We perform a DFS traverse of $Q$ in the lexicographical order and assign the indices $0,1,\ldots,q{-}1$ to the vertices of $Q$ in the order of their appearance in the traverse. Denote by $\mathit{idx}(v)$ the index of a vertex $v$. We add to our structure a bit array $M[0..q{-}1]$ initially filled with zeros. A vertex $v$ is marked iff $M[\mathit{idx}(v)] = 1$. It is easy to see that the indices of the descendants of $v$ form a contiguous segment $[\mathit{idx}(v)..j]$ for some $j \ge \mathit{idx}(v)$. So, the problem is to find for each vertex the segment of the descendant indices and then test whether there is an index $k$ in this segment such that $M[k] = 1$.

For each $v \in Q_0$, we store $\mathit{idx}(v)$ and the segment of the descendant indices explicitly using $O(\log n)$ bits. Consider a vertex $v \notin Q_0$. Let the leftmost descendant leaf of $v$ corresponds to a string $x'_{\SA[j]}$, where $j = ir{-}k$ for some $i\in [0..d/r]$ and $k \in [0..r)$. Denote by $u$ the leaf corresponding to $x'_{\SA[ir]}$. Since there are at most $2r$ vertices inserted between the leaves corresponding to $x'_{\SA[(i{-}1)r]}$ and $x'_{\SA[ir]}$ and the height of $Q$ is at most $\tau^2$, we have $0 < \mathit{idx}(u) - \mathit{idx}(v) \le 2r + \tau^2$. So, we store in $v$ the value $\mathit{idx}(u) - \mathit{idx}(v)$ using $O(\log\log n)$ bits. Obviously, one can compute $\mathit{idx}(v)$ in $O(1)$ time using $\mathit{idx}(u)$ stored explicitly. The structure occupies $O((d/r)\log n + d\log\log n) = O(d(\log\sigma + \log\log n))$ bits.

Now it is sufficient to describe how to answer the queries on the segments of the dynamic bit array $M$. We can answer the queries on the segments of length $\le \frac{\log n}2$ using a shared table occupying $O(2^{\log n/2}\log^3 n) = o(n)$ bits. So, the problem is reduced to the queries on the segments of the form $[i\log n..j\log n)$. We build a perfect binary tree $T$ with leaves corresponding to the segments $[i\log n..(i{+}1)\log n)$ for $i \in [0..q/\log n)$ (without loss of generality, we assume that $q$ is a multiple of $\log n$ and $q/\log n$ is a power of $2$). Each internal vertex $v$ of $T$ naturally corresponds to a segment $[i2^j\log n..(i{+}1)2^j\log n)$ for some $i$ and $j > 0$. Denote $c = i2^j + 2^{j-1}$. We associate with $v$ bit arrays $D_v$ and $E_v$ of lengths $2^{j{-}1}$ such that for any $k \in [1..2^{j-1}]$, $D_v[k{-}1] = 1$ iff there are ones in the segment $M[(c{-}k)\log n..c\log n{-}1]$ and, similarly, $E_v[k{-}1] = 1$ iff there are ones in $M[c\log n..(c{+}k)\log n{-}1]$. We construct on $T$ the least common ancestor structure (in the case of the perfect binary tree with $O(q/\log n)$ vertices, this can be simply done in $O(q)$ bits). Then, to answer the query on a segment $[i\log n..j\log n)$, we first find in $O(1)$ time the least common ancestor $v$ of the leaves of $T$ corresponding to the segments $[i\log n..(i{+}1)\log n)$ and $[(j{-}1)\log n..j\log n)$ and then test appropriate bits of $D_v$ and $E_v$. All in $O(1)$ time. The structure occupies $O(\frac{q}{\log n}\log\frac{q}{\log n}) = O(d)$ bits.

When we set $M[i] = 1$ for some $i \in [0..q)$, the modifications are straightforward: if the segment $[\lfloor i/\log n\rfloor..\lfloor i/\log n\rfloor{+}\log n))$ already has ones, then we are done; otherwise, for each ancestor $v$ of the leaf of $T$ corresponding to $[\lfloor i/\log n\rfloor..\lfloor i/\log n\rfloor{+}\log n)$, we scan the array $D_v$ [$E_v$] from left to right [right to left] from the appropriate position and flip all zero bits. Since there are only $O(d)$ bits in the structure, the height of $T$ is $O(\log q) = O(\log n)$, and the updates are initiated at most $q/\log n$ times, $k$ updates run in $O(d + (q/\log n)\log n + k) = O(d + k)$ time.
\qed
\end{proof}
}
\iflong
\MarkedDescProof
\fi

\noindent\mycap{Filling $\mathit{lz}$.} Denote $s_i = s[0..i]$. Let for $i\in [0..p{+}d)$, $t_i$ denotes the longest prefix of $\lvec{s_i}$ presented in $Q$. We add to each $v \in Q$ an $O(\log\log n)$-bit field $v.\mathit{mlen}$ initialized to $\tau^2$. Also, we use an integer variable $f$ that initially equals $0$.

The algorithm increases $f$ computing $|t_f|$ in each step and augments $Q$ as follows. Suppose $v \in Q$ is such that $t_{f-1}$ is a prefix of $\lab(v)$ and other vertices with this property are descendants of $v$. We say that $v$ \emph{corresponds to $t_{f-1}$}. We are to find the vertex of $Q$ corresponding to $t_f$. Suppose $p_v(s[f])$ is defined. By Lemma~\ref{BWT}, one can compute $i = \Psi(p_v(s[f]))$ in $O(1)$ time. Obviously, $x'_{\SA[i]}$ starts with $s[f]t_{f-1}$. We obtain the leaf corresponding to $x'_{\SA[i]}$ in $O(1)$ time via $L$ and $L_{\lfloor i/r\rfloor}$ and then find $w\in Q$ corresponding to $t_f$ by the WA query on the obtained leaf and the number $\min\{\tau^2, |t_{f-1}|{+}1\}$. Suppose $p_v(s[f])$ is undefined. If $v$ is the root of $Q$, then we have $|t_f| = 0$. Otherwise, we recursively process the parent $u$ of $v$ in the same way as $v$ assuming $t_{f-1} = \lab(u)$. Finally, once we have found $w\in Q$ corresponding to $t_f$, we mark the parent of $w$ using the structure of Lemma~\ref{MarkedDescendant} and assign $w.\mathit{mlen} \gets \min\{w.\mathit{mlen}, |\lab(w)|{-}|t_f|\}$.

Let $i \in [p..f{+}1]$ such that $|s[i..f{+}1]| \le \tau^2$. Suppose all positions $[0..f]$ are processed as described above. It is easy to verify that the string $s[i..f{+}1]$ has an occurrence in $s[0..f]$ iff either the vertex $v \in Q$ corresponding to $\lrange{s[i..f{+}1]}$ has a marked descendant or the parent of $v$ is marked and $|\lab(v)| - v.\mathit{mlen} \ge |s[i..f{+}1]|$. Based on this observation, the algorithm computes $\mathit{lz}$ as follows.
\begin{algorithmic}[1]
\For{$(t \gets p;\;t \le p + b;\;t \gets t + \max\{1, z\})$}
    \For{$(z \gets 0, v \gets $ the root of $Q;\;\mathbf{true};\;v \gets w, z \gets z + 1)$}
        \State increase $f$ processing $Q$ accordingly until $f = t + z - 1$
        \If{$z \ge \tau^2$} invoke the procedure of Section~\ref{SectLongFactors} to find $z$ and $\mathbf{break};$
        \EndIf
        \State find $w \in Q$ corresp. to $\lrange{s[t..t{+}z]}$ using $v$, prefix links, WA queries\label{lst:prefixFind}
        \If{$w$ is undefined}
            $\mathbf{break};$
        \EndIf
        \If{$w$ do not have marked descendants}
            \If{$\mathit{parent}(w)$ is not marked $\mathrel{\mathbf{or}} |\lab(w)|-w.\mathit{mlen} \le z$}
             $\mathbf{break};$
            \EndIf
        \EndIf
    \EndFor
    \State $\mathit{lz}[t{-}p] \gets 1;$
\EndFor
\end{algorithmic}
The lengths of the Lempel-Ziv factors are accumulated in $z$. The above observation implies the correctness. Line~\ref{lst:prefixFind} is similar to the procedure described above. Since $O(n)$ queries to the prefix links and $O(n)$ markings of vertices take $O(n)$ time, by standard arguments, one can show that the algorithm takes $O(n)$ time.

\noindent\mycap{Searching of occurrences.} Denote by $Z$ the set of all Lempel-Ziv factors of lengths $[r/2..\tau^2)$ starting in $[p..p{+}b]$. Obviously $|Z| = O(d/r)$. Using $\mathit{lz}$, we build in $O(d\log\sigma)$ time a compact trie $R$ of the strings $\{\lvec{z} \colon z\in Z\}$. We add to each $v\in R$ such that $z_v = \lrange{\lab(v)} \in Z$ the list of all starting positions of the Lempel-Ziv factors $z_v$ in $[p..p{+}b]$. Obviously, $R$ occupies $O((d/r)\log n) = O(d\log\sigma)$ bits. We construct for the strings $Z$ a succinct Aho-Corasick automaton of~\cite{Belazzougui} occupying $O((d/r)\log n) = O(d\log\sigma)$ bits.  In~\cite{Belazzougui} it is shown that the reporting states of the automaton can be associated with vertices of $R$, so that we can scan $s[0..p{+}d{-}1]$ in $O(n)$ time and store the found positions of the first occurrences of the strings $Z$ in $R$. Finally, by a DFS traverse on $R$, we obtain for each string of $Z$ the position of its first occurrence in $s[0..p{+}d{-}1]$. To find earlier occurrences of other Lempel-Ziv factors starting in $[p..p{+}b]$, we use the algorithms of Sections~\ref{SectShortFactors},~\ref{SectLongFactors}.

\section{Long Factors}\label{SectLongFactors}

\subsection{Main Tools}

Let $k\in \mathbb{N}$. A set $D \subset [0..k)$ is called a \emph{difference cover of $[0..k)$} if for any $x \in [0..k)$, there exist $y,z \in D$ such that $y - z \equiv x\pmod{k}$. Obviously $|D| \ge \sqrt{k}$. Conversely, for any $k \in \mathbb{N}$, there is a difference cover of $[0..k)$ with $O(\sqrt{k})$ elements and it can be constructed in $O(k)$ time (see \cite{BurkhardtKarkkainen}).
\iflong
\begin{example}
The set $D = \{1,2,4\}$ is a difference cover of~$[0..5)$.

\begin{minipage}{0.2\textwidth}
$$
\begin{array}{r|c|c|c|c|c}
x   & 0   & 1   & 2   & 3   & 4 \\
\hline
y,z & 1,1 & 2,1 & 1,4 & 4,1 & 1,2
\end{array}
$$
\end{minipage}%
\hfill%
\begin{minipage}{0.6\textwidth}
\includegraphics[scale=0.20]{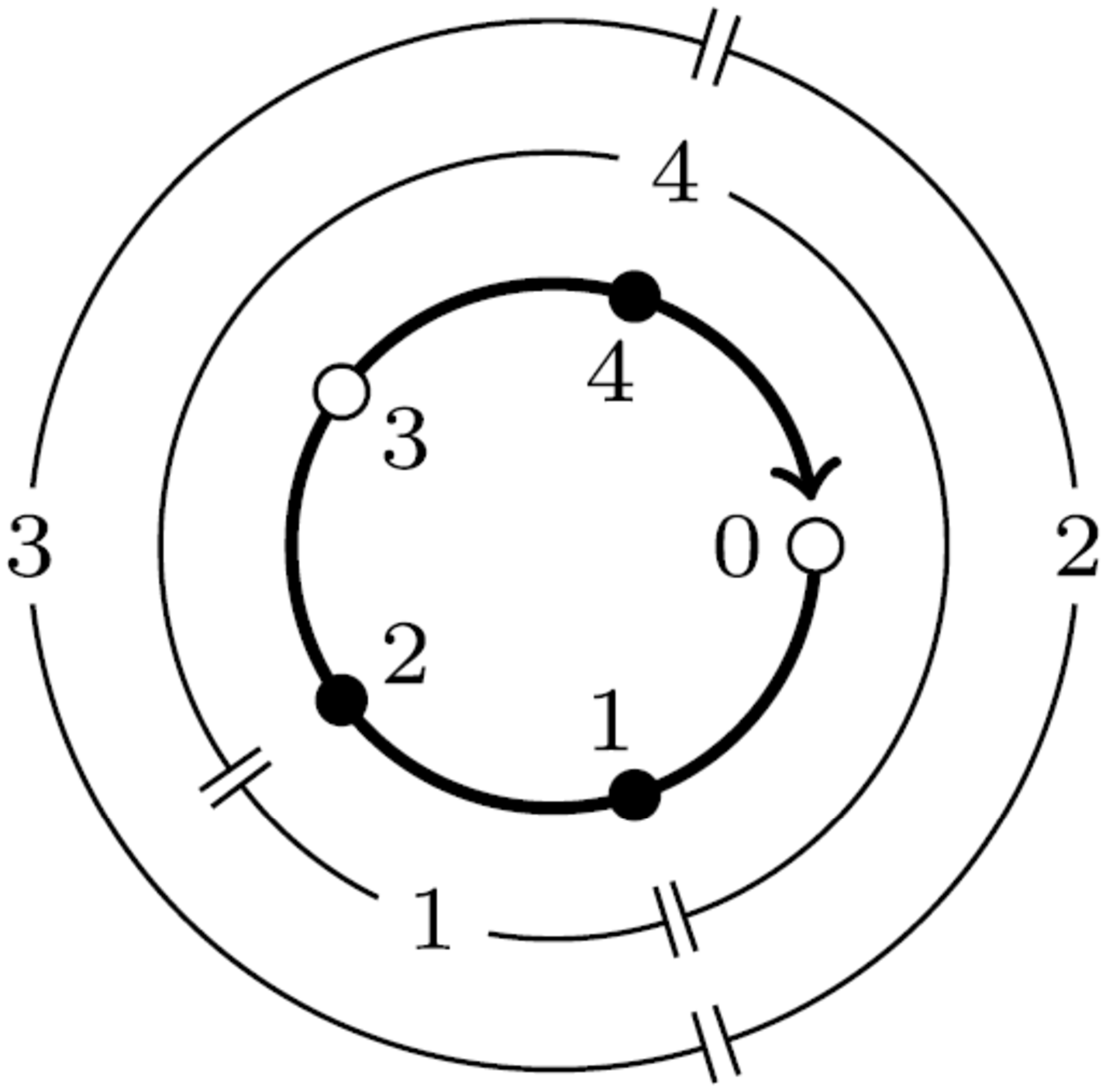}\\
\small (the figure is from~\cite{BilleGortzSachVildhoj}.)
\end{minipage}
\end{example}
\fi
\begin{lemma}[see \cite{BurkhardtKarkkainen}]
Let $D$ be a difference cover of $[0..k)$. For any integers $i,j$, there exists $d \in [0..k)$ such that $(i - d) \bmod k \in D$ and $(j - d) \bmod k \in D$.\label{DiffCoverProperty}
\end{lemma}

An \emph{ordered tree} is a tree whose leaves are totally ordered (e.g, a trie).
\begin{lemma}[see~\cite{NavarroSadakane}]
In $O(k\log k)$ bits of space we can maintain an ordered tree with at most $k$ vertices under the following operations:\\
1.insertion of a new leaf (possibly splitting an edge) in $O(\log k)$ time;\\
2.searching of the leftmost/rightmost descendant leaf of a vertex in $O(\log k)$ time.%
\label{OrderedTree}
\end{lemma}

\begin{lemma}[see \cite{BenderColeDemaineFarachColtonZito}]
A linked list can be designed to support the following operations:
1. insertion of a new element in $O(1)$ amortized time;
2. determine whether $x$ precedes $y$ for given elements $x$ and $y$ in $O(1)$ time.
\label{OrderedList}
\end{lemma}

\iflong
To support fast navigation in tries, we associate with each vertex $v$ a dictionary mapping the first letters in the labels written on the outgoing edges of $v$ to the corresponding children of $v$. So, whether a trie contains a string with a prefix $w$ can be checked in $O(|w|\log\rho)$ time, where $\rho$ is the alphabet size. Notice that a compact trie for a set of $k$ substrings of the string $s$ can be stored in $O(k\log n)$ bits using pointers for the edge labels. But the described searching time is too slow for our purposes, so, using packed strings and fast string dictionaries, we improve our tries with the operations provided in the following lemma.
\fi
\iflong
\begin{lemma}
\else
\begin{lemma}[see arXiv:1504.06712]
\fi
In $O(k\log n)$ bits of space we can maintain a compact trie for at most $k$ substrings of $s$ under the following operations:\\
1. insertion of a string $w$ in $O(|w|/r + \log n)$ amortized time;\\
2. searching of a string $w$ in $O(|u|/r + \log n)$ time, where $u$ is the longest prefix of $w$ present in the trie; we scan $w$ from left to right $r$ letters at a time and report the vertices of the trie corresponding to the prefixes of lengths $r, 2r, \ldots, \lfloor|u|/r\rfloor r$, and $|u|$ immediately after reading these prefixes.%
\label{Trie}
\end{lemma}
\newcommand{\TrieProof}{
\begin{figure}
\center
\includegraphics[scale=0.55]{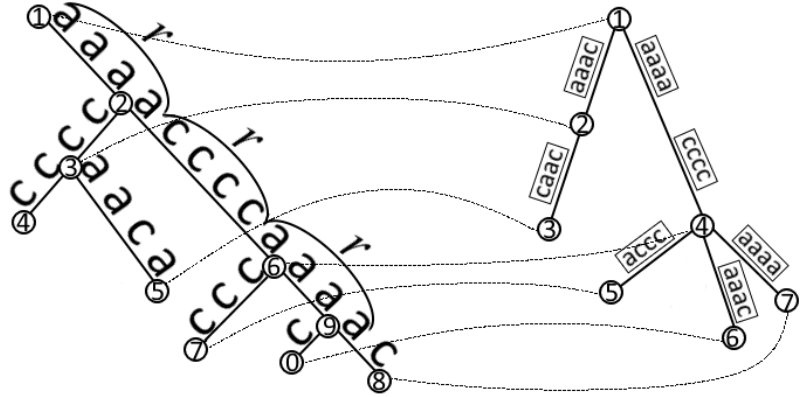}
\caption{\small A compact trie $T$ is on the left; the corresponding ternary tree $T'$ is on the right. If $r = 4$, the searching of $w = aaaaccccaaaac$ reports the vertices 6,6,8,8 corresponding to the prefixes of lengths $r$, $2r$, $3r$, and $|w|$, respectively.}
\label{fig:trie}
\end{figure}
\begin{proof}
Denote by $S$ the set of all strings stored in $T$. For a substring $t$ of the string $s$, denote by $t'$ a string of length $\lfloor|t|/r\rfloor$ such that for any $i \in [0..|t'|)$, $t'[i]$ is equal to the packed string $t[ri..r(i{+}1){-}1]$. We maintain a special compact trie $T'$ containing the set of strings $\{t' \colon t \in S\}$: the dictionaries associated with the vertices of $T'$ are organized in such a way that the searching and insertion of a string $w'$ both work in $O(|w'| + \log k)$ amortized time; such tries are called \emph{dynamic ternary trees} (see \cite{FranceschiniGrossi} for a comprehensive list of references). For each $v\in T$, we insert in $T'$ a vertex corresponding to the string $t'$ (if there is no such vertex), where $t = \lab(v)$ (consider the vertices~3 on the left and~2 on the right of Fig.~\ref{fig:trie}). All vertices of $T'$ are augmented with the pointers to the corresponding vertices of $T$ (depicted as dashed lines in Fig.~\ref{fig:trie}).

Let $w$ be a string to be searched in $T$. Using the pointers of $T'$, we can report vertices corresponding to the prefixes $w[0..r{-}1]$, $w[0..2r{-}1], \ldots, w[0..|w'|r{-}1]$ while traverse $T'$. Denote by $u$ the longest prefix of $w$ presented in $T$. Once $u'$ is found in $T'$ in $O(|u'| + \log k)$ time, we start to traverse $T$ reading the string $u[|u'|r..|u|{-}1]$ from the position corresponding to $u[0..|u'|r{-}1]$. This operation requires additional $O(r\log\sigma) = O(\log n)$ time. The insertion is analogous.\qed
\end{proof}
}
\iflong
\TrieProof
\fi

In the \emph{dynamic tree range reporting problem} one has ordered trees $T_1$ and $T_2$ and a set of pairs $Z = \{(x_1^i,x_2^i)\}$, where $x_1^i$ and $x_2^i$ are leaves of $T_1$ and $T_2$, respectively (see Fig.~\ref{fig:treerange}); the query asks, for given vertices $v_1\in T_1$ and $v_2\in T_2$, to find a pair $(x_1,x_2) \in Z$ such that $x_1$ and $x_2$ are descendants of $v_1$ and $v_2$, respectively; the update inserts new pairs in $Z$ or new vertices in $T_1$ and $T_2$. To solve this problem, we apply the structure of~\cite{Blelloch} and Lemmas~\ref{OrderedTree} and~\ref{OrderedList}.
\iflong
\begin{lemma}
\else
\begin{lemma}[see arXiv:1504.06712]
\fi
The dynamic tree range reporting problem with $|Z|\le k$ can be solved in $O(k\log k)$ bits of space with updates and queries working in $O(\log k)$ amortized time. \label{OrthogonalTree}
\end{lemma}
\newcommand{\OrthTreeProof}{
\begin{proof}
To prove this Lemma, we need an additional tool. In the \emph{dynamic orthogonal range reporting problem} one has two linked lists $X$ and $Y$, and a set of pairs $Z = \{(x_i,y_i)\}$, where $x_i \in X$ and $y_i \in Y$; the query asks to report for given elements $x_1, x_2 \in X$ and $y_1,y_2 \in Y$, a pair $(x,y) \in Z$ such that $x$ lies between $x_1$ and $x_2$ in $X$, and $y$ lies between $y_1$ and $y_2$ in $Y$; the update inserts new pairs in $Z$ or new elements in $X$ or $Y$.
\begin{lemma}[see~\cite{Blelloch}]
The dynamic orthogonal range reporting problem on at most $k$ pairs can be solved in $O(k\log k)$ bits of space with updates and queries working in $O(\log k)$ amortized time.
\label{OrthogonalRange}
\end{lemma}
We maintain the ordered tree structure of Lemma~\ref{OrderedTree} on $T_1$ and $T_2$. The order on the lists of leaves of $T_1$ and $T_2$ is maintained with the aid of enhanced linked lists of Lemma~\ref{OrderedList}. To process queries efficiently, we build the dynamic orthogonal range reporting structure of Lemma~\ref{OrthogonalRange} on these lists and the set of pairs $Z$. These structures take overall $O(k\log k)$ bits of space. By Lemmas~\ref{OrderedList},~\ref{OrderedTree},~\ref{OrthogonalRange}, the update of $T_1$, $T_2$, or $Z$ requires $O(\log k)$ amortized time.

Suppose we process a query for vertices $v_1\in T_1$ and $v_2\in T_2$. We obtain the leftmost and rightmost descendant leaves of $v_1$ and $v_2$ using Lemma~\ref{OrderedTree}. Then we report a desired pair from $Z$ (or decide that there are no such pairs) using Lemma~\ref{OrthogonalRange}. By Lemmas~\ref{OrderedTree} and \ref{OrthogonalRange}, the query takes $O(\log k)$ amortized time.\qed
\end{proof}
}
\iflong
\OrthTreeProof
\fi

\subsection{Algorithm for Long Factors}

\noindent\mycap{Data structures.} At the beginning, using the algorithm of~\cite{BurkhardtKarkkainen}, our algorithm constructs a difference cover $D$ of $[0..\tau^2)$ such that $|D| = \Theta(\tau)$. Denote $M = \{i\in[0..n) \colon i\bmod\tau^2 \in D\}$. The set $M$ is the basic component in our constructions.

Suppose the algorithm has reported the Lempel-Ziv factors $z_1, z_2, \ldots, z_{k-1}$ and already decided that $|z_k| \ge \tau^2$ applying the procedure of Section~\ref{SectMidFactors}. Denote $p = |z_1z_2\cdots z_{k-1}|$. We use an integer variable $z$ to compute the length of $|z_k|$ and $z$ is initially equal to $\tau^2$. Let us first discuss the related data structures.

We use an auxiliary variable $t$ such that $p \le t < p + z$ at any time of the work; initially $t = p$. Denote $s_i = s[0..i]$. Our main data structures are compact tries $S$ and $T$: $S$ contains the strings $\lvec{s_i}$ and $T$ contains the strings $s[i{+}1..i{+}\tau^2]$ for all $i\in[0..t) \cap M$ (we append $\tau^2$ letters $s[0]$ to the right of $s$ so that $s[i{+}1..i{+}\tau^2]$ is always defined). Both $S$ and $T$ are augmented with the structures supporting the searching of Lemma~\ref{Trie} and the tree range queries of Lemma~\ref{OrthogonalTree} on pairs of leaves of $S$ and $T$. Since $s[0]$ is a sentinel letter, each $\lvec{s_i}$, for $i \in [0..t) \cap M$, is represented in $S$ by a leaf. The set of pairs for our tree range reporting structure contains the pairs of leaves corresponding to $\lvec{s_i}$ in $S$ and $s[i{+}1..i{+}\tau^2]$ in $T$ for all $i \in [0..t) \cap M$ (see Fig.~\ref{fig:treerange}). Also, we add to $S$ the WA structure of Lemma~\ref{WeightedAncestor}.%
\begin{figure}
\center
\includegraphics[scale=0.35]{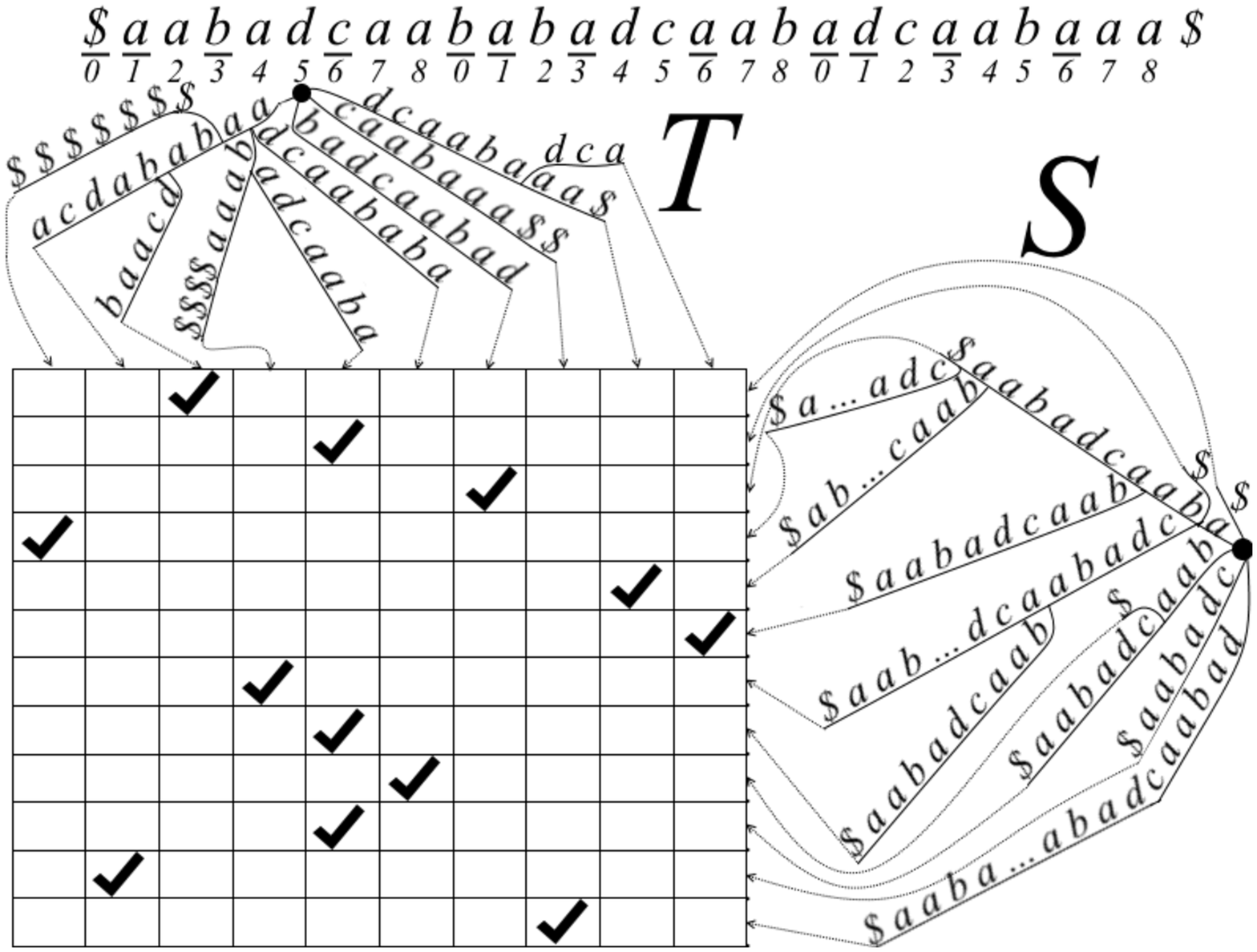}
\caption{\small $\tau = 3$, $D = \{0, 1, 3, 6\}$ is a diff. cover of $[0..\tau^2)$, positions in $M$ are underlined.}
\label{fig:treerange}
\end{figure}

Let us consider vertices $v \in S$ and $v' \in T$ corresponding to strings $\lvec{t}_v$ and $t_{v'}$, respectively. Denote by $\mathrm{treeRng}(v, v')$ the tree range query that returns either $\mathbf{nil}$ or a suitable pair of descendant leaves of $v$ and $v'$. We have $\mathrm{treeRng}(v, v') \ne \mathbf{nil}$ iff there is $i \in [0..t) \cap M$ such that $s[i{-}|t_v|{+}1..i]s[i{+}1..i{+}|t_{v'}|] = \lvec{t}_vt_{v'}$.

Since $|M| \le \frac{n}{\tau^2}|D| = O(\frac{n}{\tau})$, it follows from Lemmas~\ref{WeightedAncestor},~\ref{Trie},~\ref{OrthogonalTree} that $S$ and $T$ with all related structures occupy at most $O(\frac{n}{\tau} \log n) = O(\epsilon n)$ bits.

\noindent\mycap{The algorithm.} Suppose the factor $z_k$ occurs in a position $x \in [0..p)$; then, by Lemma~\ref{DiffCoverProperty}, there is a $d \in [0..\tau^2)$ such that $x + |z_k| - d \in M$ and $p + |z_k| - d \in M$. Based on this observation, our algorithm, for each $t \in M \cap [p..z)$, finds the vertex $v\in S$ corresponding to $\lrange{s[p..t]}$ and the vertex $v'\in T$ corresponding to as long as possible prefix of $s[t{+}1..n{+}\tau^2]$ such that $\mathrm{treeRng}(v, v') \ne \mathbf{nil}$ and with the aid of this bidirectional search, we further increase $z$ if it is possible.
\begin{algorithmic}[1]
\For{$(t \gets \min\{i \ge p \colon i \in M\};\;t < p + z;\;t \gets \min\{i > t \colon i \in M\})$}
    \State $x \gets$ the length of the longest prefix of $s[t{+}1..t{+}\tau^2]$ present in $T$\label{lst:longestPrefixT}
    \State $y \gets$ the length of the longest prefix of $\lvec{s_t}$ present in $S$\label{lst:longestPrefixS}%
    \If{$y < t - p + 1$}
        go to line~\ref{lst:insertTrie}
    \EndIf
    \State $v \gets $ the vertex corresp. to the longest prefix of $\lvec{s_t}$ present in $S$\label{lst:findPrefix}
    \State $v \gets \mathrm{weiAnc}(v, t - p + 1);$\label{lst:wancestor}
    \For{$j = t, t{+}r, t{+}2r, \ldots, t{+}\lfloor x/r\rfloor r, x$ and $v' \in T$ corresp. to $s[t{+}1..j]$}\label{lst:loopTraverse}%
        \If{$j \ge p + z$} \Comment{$|s[p..j]| > |s[p..p{+}z{-}1]|$}
            \If{$\mathrm{treeRng}(v, v') = \mathbf{nil}$} \label{lst:treeRng}
                \State $j \gets \max\{j'{\colon}\mathrm{treeRng}(v, u){\ne}\mathbf{nil}$ for $u\in T$ corresp. $s[t{+}1..j']\};$\label{lst:binary}%
            \EndIf
            \State $z \gets \max\{z, j - p + 1\};$\label{lst:increaseZ}
            \If{$\mathrm{treeRng}(v, v') = \mathbf{nil}$}
                $\mathbf{break};$
            \EndIf
        \EndIf
    \EndFor \label{lst:loopTraverseEnd}
    \State insert $s[t{+}1..t{+}\tau^2]$ in $T$, $\lvec{s_t}$ in $S$; process the pair of the corresp. leaves\label{lst:insertTrie}
\EndFor
\end{algorithmic}
Some lines need further clarification. Here $\mathrm{weiAnc}(v, i)$ denotes the WA query that returns either the ancestor of $v$ with the minimal weight $\ge i$ or $\mathbf{nil}$ if there is no such ancestor; we assume that any vertex is an ancestor of itself. Since $M$ has period $\tau^2$, one can compute, for any $t$, $\min\{i > t \colon i\in M\}$ in $O(1)$ time using an array of length $\tau^2$ for example. The operations on $T$ in lines~\ref{lst:longestPrefixT},~\ref{lst:insertTrie} take, by Lemma~\ref{Trie}, $O(\tau^2/r + \log n)$ time. To perform the similar operations on $S$ in lines~\ref{lst:longestPrefixS},~\ref{lst:findPrefix},~\ref{lst:insertTrie}, we use other techniques (discussed below) working in the same time. The loop in line~\ref{lst:loopTraverse} executes exactly the procedure described in Lemma~\ref{Trie}. To compute $j$ in line~\ref{lst:binary}, we perform the binary search on at most $r$ ancestors of the vertex $v'$; thus, we invoke $\mathrm{treeRng}$ $O(\log r)$ times in line~\ref{lst:binary}.

Let us prove the correctness. Suppose we have $\tau^2 \le z < |z_k|$ in some iteration. It suffices to show that the algorithm cannot terminate with this value of $z$. Let $z_k$ occur in a position $x \in [0..p)$. By Lemma~\ref{DiffCoverProperty}, there is a $d \in [0..\tau^2)$ such that $x + z - d \in M$ and $p + z - d \in M$. Thus, the string $s[p..p{+}z{-}d]$ is presented in $S$ when $t = p + z - d$ and we find the corresponding vertex $v$ in line~\ref{lst:wancestor}. Moreover, the string $s[p{+}z{-}d{+}1..p{+}z]$ is presented in $T$ and we find the vertex corresponding to this or a longer string in the loop~\ref{lst:loopTraverse}--\ref{lst:loopTraverseEnd}. Denote this vertex by $w$; $w$ is either $v'$ or $u$ in line~\ref{lst:binary}. Obviously, $\mathrm{treeRng}(v, w) \ne \mathbf{nil}$, so, we increase $z$ in line~\ref{lst:increaseZ}.

Let us estimate the running time. The main loop performs $O(|z_k|/\tau)$ iterations. The operations in lines~\ref{lst:longestPrefixT}, \ref{lst:longestPrefixS}, \ref{lst:findPrefix}, \ref{lst:insertTrie} require, as mentioned above, $O(\tau^2/r + \log n)$ time (some of them will be discussed in the sequel). One WA query and one modification of the tree range reporting structure take, by Lemmas~\ref{WeightedAncestor} and~\ref{OrthogonalTree}, $O(\log n)$ time. By Lemma~\ref{Trie}, the traverse of $T$ in line~\ref{lst:loopTraverse} requires $O(\tau^2/r + \log n)$ time. For each fixed $t$, every time we perform $\mathrm{treeRng}$ query in line~\ref{lst:treeRng}, except probably for the first and last queries, we increase $z$ by $r$. Hence, the algorithm executes at most $O(|z_k|/\tau + |z_k|/r)$ such queries in total. Finally, in line~\ref{lst:binary} we invoke $\mathrm{treeRng}$ at most $O(\log r)$ times for every fixed $t$. Putting everything together, we obtain $O(\frac{|z_k|}{\tau}(\tau^2/r + \log n) + \frac{|z_k|}{r}\log n + \frac{|z_k|}{\tau}\log r \log n) = O(|z_k|\log\sigma + |z_k|\log r) = O(|z_k|(\log\sigma + \log\log n))$ overall time.

One can find the position of an early occurrence of $z_k$ from the pairs of leaves reported in lines~\ref{lst:treeRng},~\ref{lst:binary}. Now let us discuss how to insert and search strings in $S$.

\noindent\mycap{Operations on $S$.} The operations on $S$ are based on the fact that for any $i \in [\tau^2..n) \cap M$, $i - \tau^2 \in M$. Let $u$ and $v$ be leaves of $S$ corresponding to some $\lvec{s_j}$ and $\lvec{s_k}$. To compare $\lvec{s_j}$ and $\lvec{s_k}$ in $O(1)$ time via $u$ and $v$, we store all leaves of $S$ in a linked list $K$ of Lemma~\ref{OrderedList} in the lexicographical order. To calculate $\lcp(\lvec{s_j}, \lvec{s_k})$ in $O(\log n)$ time via $u$ and $v$, we put all leaves of $S$ in an augmented search tree $B$. Finally, we augment $S$ with the ordered tree structure of Lemma~\ref{OrderedTree}.

Denote $s'_i = \lrange{s[i{-}\tau^2{+}1..i]}$. We add to $S$ a compact trie $S'$ containing $s'_i$ for all $i \in [0..t) \cap M$ (we assume $s[0]{=}s[-1]{=}\ldots$, so, $S'$ is well-defined). The vertices of $S'$ are linked to the respective vertices of $S$. Let $w$ be a leaf of $S'$ corresponding to a string $s'_i$. We add to $w$ the set $H_w = \{(p^j_1, p^j_2) \colon j\in [0..t)\cap M\text{ and }s'_j = s'_i\}$, where $p^j_1$ and $p^j_2$ are the pointers to the leaves of $S$ corresponding to $\lvec{s}_{j{-}\tau^2}$ and $\lvec{s_j}$, respectively; $H_w$ is stored in a search tree in the lexicographical order of the strings $\lvec{s}_{j{-}\tau^2}$ referred by $p^j_1$, so, one can find, for any $k \in [0..t{+}\tau^2) \cap M$, the predecessor or successor of the string $\lvec{s}_{k-\tau^2}$ in $H_w$ in $O(\log n)$ time. It is straightforward that all these structures occupy $O(\frac{n}{\tau}\log n) = O(\epsilon n)$ bits.

Suppose $S$ contains $\lvec{s_i}$ for all $i \in [0..t)\cap M$ and we insert $\lvec{s_t}$. We first search $s'_t$ in $S'$. Suppose $S'$ does not contain $s'_t$. We insert $s'_t$ in $S'$ in $O(\tau^2/r + \log n)$ time, by Lemma~\ref{Trie}, then add to $S$ the vertices corresponding to the new vertices of $S'$ and link them to each other. Using the structure of Lemma~\ref{OrderedTree} on $S$, we find the position of $\lvec{s_t}$ in $K$ in $O(\log n)$ time. All other structures are easily modified in $O(\log n)$ time. Now suppose $S'$ has a vertex $w$ corresponding to $s'_t$. In $O(\log n)$ time we find in $H_w$ the pairs $(p^{j}_1, p^{k}_2)$ and $(p^{j}_1, p^{k}_2)$ such that $p^{j}_1$ points to the predecessor $\lvec{s}_{j{-}\tau^2}$ of $\lvec{s}_{t{-}\tau^2}$ in $H_w$ and $p^{k}_1$ points to the successor $\lvec{s}_{k{-}\tau^2}$. So, the leaf corresponding to $\lvec{s_t}$ must be between $\lvec{s}_{j}$ and $\lvec{s}_{k}$. Using $B$, we calculate $\lcp(\lvec{s}_{j}, \lvec{s_t}) = \lcp(\lvec{s}_{j{-}\tau^2}, \lvec{s}_{t{-}\tau^2}) + \tau^2$ and, similarly, $\lcp(\lvec{s}_{k}, \lvec{s_t})$ in $O(\log n)$ time and then find the position where to insert the new leaf by WA queries on $S$. All other structures are simply modified in $O(\log n)$ time. Thus, the insertion takes $O(\tau^2/r + \log n)$ time. One can use a similar algorithm for the searching of $\lvec{s_t}$.

\bibliographystyle{splncs03}
\bibliography{lz}

\iflong
\fi
\end{document}